\documentstyle[twoside,fleqn,espcrc2,epsf]{article}

\epsfverbosetrue
 
\newcommand{\AmS}{{\protect\the\textfont2
  A\kern-.1667em\lower.5ex\hbox{M}\kern-.125emS}}

\title{The Charmonium Spectrum on the Lattice: A Status Report%
\thanks{presented at {\em Lattice'92}, Amsterdam, Sep. 15-19, 1992}}

\author{Aida X. El-Khadra \\
\medskip
Theory Group, Fermilab, P.O. Box 500, Batavia Il 60510}
\begin{document}
 
\begin{abstract}
We present our most recent results on the charmonium spectrum using
relativistic Wilson fermions. We study the dependence of the spectrum 
on the charm quark mass and the Wohlert-Sheikholeslami improvement term.
\end{abstract}
 
\maketitle
 
\section{INTRODUCTION}
This is a status report on our calculation of the charmonium spectrum.
Our first results were presented last year \cite{lat91}.
An important aspect of this calculation, the determination of $\alpha_s$ 
from the charmonium  spectrum, has already been reported separately \cite{prl}.
The detailed results of this study will be presented in a future 
publication \cite{future}.

The calculation of experimentally well measured quantities in the charmonium 
spectrum serves several purposes.
The investigation of physical quantities
that are sensitive to the systematic errors present in 
lattice calculations can lead to a better understanding and possibly
the removal of these errors.
On the other hand, a quantity that is relatively independent of the
systematic errors of the lattice calculation can be used to obtain information
about fundamental parameters of the Standard Model.

In Refs.~\cite{lat91,prl} we argued that the 1P-1S splitting falls into the latter class
of quantities and therefore lends itself to an accurate determination of the
strong coupling constant of QCD.
The splitting between the $J/\psi$ and $\eta_c$, on the other hand, 
is expected to strongly depend on the leading lattice spacing errors. 
This splitting and the leptonic decay amplitude of the $J/\psi$ are expected 
to also show some dependence on the charm quark mass. One might therefore
use these two quantities in a study of the former type.

Some of the details of the calculation are described in section~\ref{sec:calc}.
The effect of the omission of sea quarks has already been discussed in detail in our
analysis of $\alpha_s$. In section~\ref{sec:sea} we discuss their effect on the 
hyperfine splitting and the leptonic decay amplitude of the $J/\psi$.
The results of our study of the dependence of the spectrum on the charm quark mass 
and ${\cal O}(a)$ lattice spacing errors are presented in section~\ref{sec:res}.


\section{THE CALCULATION} \label{sec:calc}

We analyze three different lattices ($12^3\times 24$, $16^3\times 32$, $24^4$)
at three different couplings ($\beta = 5.7, 5.9, 6.1$) such that the volumes 
are similar while the lattice spacing varies by about a factor of two
(see Ref.~\cite{lat91,prl} for more details). 
In order to investigate the effects of lattice spacing errors on the spectrum,
we use Wilson fermions with ($c = 1.4$) and without ($c = 0$) the
${\cal O} (a)$ correction term \cite{ws}. 
We study the charm quark mass dependence by varying the charm hopping parameter,
$\kappa_{\rm charm}$, over a reasonable range for each lattice and 
fermion action (or choice of $c$).


The 2-pt. functions used for the mass splittings have already been
described in detail in Ref.~\cite{lat91}. The mesons are created and destroyed
by operators,
$\chi (x) = \sum_{\rm r} \psi(x+{\rm r}) \Gamma \bar{\psi} (x) S({\rm r})$,
with exponential spreading functions, $S({\rm r})$, that optimize the overlap
with the ground states.
The leptonic matrix element for the $J/\psi$ is extracted from the 2-pt. function
($\Gamma = \gamma_i$)
\begin{equation} \label{eq:GV}
G_2^V (t;\mu, i) = \sum_{\rm x} \langle V_{\mu} (x) \chi^{\dagger}_i (0) \rangle \;\;.
\end{equation}
\mbox{$V_{\mu} (x) = \bar{\psi}(x) \gamma_{\mu} \psi (x)$} is the local vector current.
With the convention \mbox{$\langle p | p' \rangle = (2 \pi)^3 \delta^3(p-p')$}, 
Eq.~(\ref{eq:GV}) takes the asymptotic form
\begin{equation}
G_2^V (t;\mu, i) \rightarrow \langle 0|V_{\mu}|J/\psi \rangle 
  \langle J/\psi | \chi^{\dagger} |0 \rangle e^{-m t} \;\;.
\end{equation}
We parametrize the matrix element in terms of $V_{\psi}$ as
\begin{equation}  \label{eq:vpsi}
\langle 0 | V_{\mu} | J/\psi \rangle = \epsilon_{\mu} V_{\psi} \;\;.
\end{equation} 
$V_{\psi}$ is related to the decay constant $f^{-1}_{\psi}$ used by 
other groups \cite{ukqcd} by $V_{\psi} = (m^3_{\psi}/2)^{1/2} f^{-1}_{\psi}$.
We renormalize the current non-perturbatively with the charge matrix element
$\langle J/\psi | V_4 | J/\psi \rangle = Z_V^{-1}$. As remarked in Refs.~\cite{ask,pbm},
the perturbative calculation of the local current renormalization $Z_V$ 
(using improved perturbation theory \cite{pp}) reproduces 
the non-perturbative result to  a few $\%$ in the case of Wilson ($c=0$) fermions,
if the quark fields are properly normalized.

\section{THE EFFECT OF SEA QUARKS}   \label{sec:sea}

A nonrelativistic system, like charmonium, is well described by a potential.
We can therefore understand the dominant effects of the sea quarks on the spectrum
via their effect on the potential. 
Following the arguments of Ref.~\cite{prl}, in setting the scale $a^{-1}$ with the
1P-1S splitting, we require in effect the potentials of the full and the quenched
theory to match at the relevant physics scale which is the intermediate distance
scale of $\sim 400 - 750$ MeV. 
We then expect the (perturbative) short distance potential to be too weak,
as a result of the zero flavor $\beta$-function being slightly too large.

For quantities that are dominated by short distance effects, like the hyperfine
splitting and leptonic decay amplitude of the $J/\psi$, we have to consider the
sea quark effects on the short distance potential.
This is best investigated within the context of a potential model.
One expects for the hyperfine splitting to lowest order \cite{quigg}
\begin{equation}
\Delta m (J/\psi - \eta_c) \sim \frac{\alpha_s(m_c)}{m_c^2} |\Psi(0)|^2 \;\;.
\end{equation}
Similarly, according to the van Royen - Weisskopf formula we expect for the leptonic
matrix element
\begin{equation}
V_{\psi} \sim |\Psi (0)| \;\;.
\end{equation}
Thus, the effect of the sea quarks on the hyperfine splitting
and the leptonic matrix element can be estimated by their effect on the 
wave function at the origin, $\Psi(0)$, in a potential model.

\begin{figure}
\epsfxsize=0.49\textwidth
\epsfbox{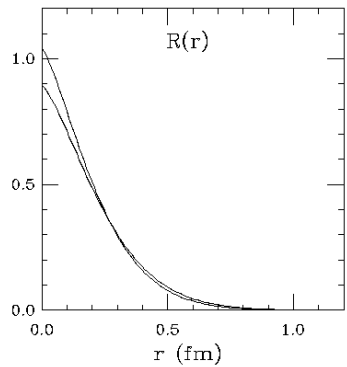}
\caption{The radial wave function for the 1S state in the Richardson potential model.
The solid line is for $n_f = 3$ and the dashed line for $n_f = 0$.}  \label{fig:richwf}
\end{figure}

The Richardson potential is convenient for our purposes, because it
incorporates asymptotic freedom and confinement in a simple form \cite{rich}:
\begin{equation}
V(q^2) = C_F \frac{4\pi}{\beta^{(n_f)}_0} \frac{1}{q^2 \ln{(1+q^2/\Lambda^2)}} \;\;,
\end{equation}
with $\beta_0^{(n_f)} = 11 - 2 n_f /3$.
Figure~\ref{fig:richwf} shows the wave functions of the 1S state, obtained from
fitting the Richardson potential to the experimental charmonium spectrum
with $n_f = 3$ and $n_f = 0$ in comparison. We find
for the ratio of wave functions at the origin:
\begin{equation} \label{eq:wfcor}
\frac{\Psi^{(0)}(0)}{\Psi^{(3)}(0)} = 0.86 \;\;.
\end{equation}

The hyperfine splitting receives an additional correction from $\alpha_s(m_c)$.
The calculation of the effects of the sea quarks on $\alpha_s(\pi/a)$ is 
described in detail in Ref.~\cite{prl}. The resulting correction for $\alpha_s(m_c)$ 
is analogously:
\begin{equation}
\frac{\alpha_s^{(0)} (m_c)}{\alpha_s^{(3)} (m_c)} = 0.81 \pm 0.06 \;\;.
\end{equation}
In summary, we estimate the quenched hyperfine splitting
to be reduced by $40\,\%$ from the experimental value, 
$\Delta m (J/\psi - \eta_c)^{\rm exp} = 117.3$ MeV to
\begin{equation}  \label{eq:hypquen}
\Delta m (J/\psi - \eta_c)^{\rm quenched} \simeq 70 \; {\rm MeV} \;\;.
\end{equation}
The quenched leptonic matrix element just recieves the correction 
in Eq.~(\ref{eq:wfcor}), a $14\,\%$ reduction from the experimental result 
$V_{\psi}^{\rm exp} = 0.509$ GeV$^{3/2}$:
\begin{equation} \label{eq:Vquen}
V_{\psi}^{\rm quenched} = 0.438 \; {\rm GeV}^{3/2} \;\;.
\end{equation}

There is certainly no reason to believe that the sea quarks do not
change the potential (non-perturbatively) in a way that has not been taken
into account here. Eqs.~(\ref{eq:hypquen}) and (\ref{eq:Vquen}) are therefore
only rough estimates of the possible effect. However, from the above considerations
it should be expected that the omission of sea quarks reduces both the hyperfine
splitting and the leptonic width from their experimental values.

\section{RESULTS} \label{sec:res}

\begin{figure}
\epsfxsize=0.49\textwidth
\epsfbox{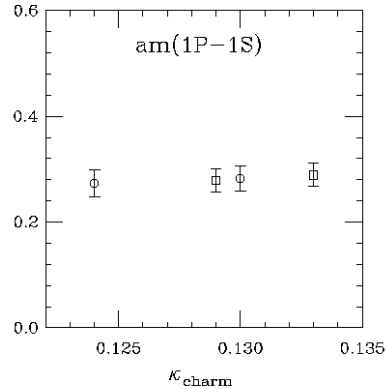}
\caption{The 1P-1S splitting in lattice units vs. $\kappa_{\rm charm}$ on the
$16^3\times32$, $\beta = 5.9$ lattice; the circles are for $c = 1.4$, the
squares are for $c=0$.} \label{fig:1p1s}
\end{figure}

\begin{figure}
\epsfxsize=0.49\textwidth
\epsfbox{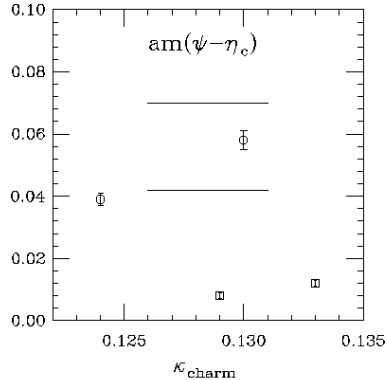}
\caption{The $J/\psi$-$\eta_c$ splitting in lattice units vs. 
$\kappa_{\rm charm}$ on the $16^3\times32$, $\beta = 5.9$ lattice; 
the circles are for $c = 1.4$, the squares are for $c=0$. The dashed line is the
experimnetal splitting; the solid line is the splitting of Eq.~(9).} 
\label{fig:mhyp}
\end{figure}

\begin{figure}
\epsfxsize=0.49\textwidth
\epsfbox{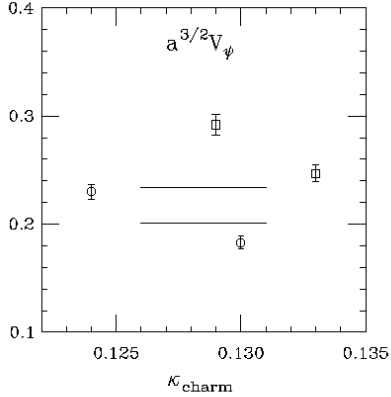}
\caption{The leptonic matrix element, $V_{\psi}$ in lattice units, vs. 
$\kappa_{\rm charm}$ on the $16^3\times32$, $\beta = 5.9$ lattice; 
the circles are for $c = 1.4$, the squares are for $c=0$. The dashed line is the
experimental value; the solid line is the value of Eq.~(10).} \label{fig:Vpsi}
\end{figure}

Figures~\ref{fig:1p1s}, \ref{fig:mhyp} and \ref{fig:Vpsi} show the
1P-1S splitting, the $J/\psi$ - $\eta_c$ splitting, and the leptonic
matrix element of the $J/\psi$ as functions of the charm hopping parameter
for Wilson fermions with ($c=1.4$) and without ($c=0$) the improvement term
on the $16^3\times32$ ($\beta = 5.9$) lattice as representative examples 
of our results on all three lattices. 
As expected, we find the dependence of the 1P-1S splitting on the charm
quark mass (parametrized by $\kappa_{\rm charm}$) and the coefficient 
of the improvement term, $c$, to be very
weak, smaller than our statistical errrors.
This confirms our previous arguments \cite{lat91,prl} to use this splitting
to determine the scale $a^{-1}$ in lattice calculations and subsequently
extract the strong coupling constant from it. Our previous result
for $\alpha_s$ remains unchanged after this study.

\begin{figure}
\epsfxsize=0.49\textwidth
\epsfbox{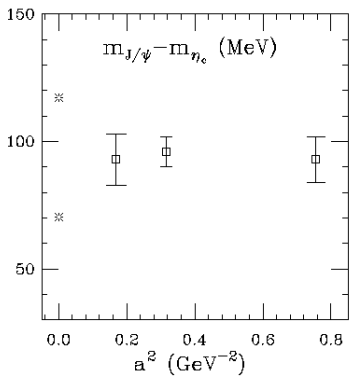}
\caption{The $J/\psi$-$\eta_c$ splitting vs. $a^2$.
The two points on the left (burst) represent the experimental splitting (upper)
and the expectation for the quenched result (lower).} \label{fig:hyp}
\end{figure}

Both the hyperfine splitting as well as the leptonic matrix element vary
significantly with the two action parameters, $\kappa_{\rm charm}$ and $c$.
However, from the considerations in section~\ref{sec:sea}, it is clear
that a phenomenological determination of these parameters is not possible within 
the quenched approximation without a better understanding of the effects of the 
sea quarks on the static potential.
We therefore take the mean field value $c=1.4$ as our best estimate
for the coefficient of the improvement term.
Setting the charm quark mass with the leptonic matrix element, $V_{\psi}$
(using Eq.~(\ref{eq:Vquen})), we extract hyperfine splittings as shown in
figure~\ref{fig:hyp} for all three lattices. The splitting in physical
units is graphed as a function of $a^2$ for better visiblility. Also shown
are the experimental and the expected quenched values. 
We find very little variation of the hyperfine splitting with the lattice
spacing, indicating small residual lattice spacing errors. Our final
result 
\begin{equation}
\Delta m(J/\psi - \eta_c) = 93 \pm 10 \; {\rm MeV}
\end{equation}
(statistical error only) lies within the expectations for a 
quenched calculation, slightly below the experimentally measured splitting.


\section*{ACKNOWLEDGEMENTS}
I thank G. Hockney, A. Kronfeld and P. Mackenzie for
an enjoyable collaboration. This calculation was performed on the 
Fermilab lattice supercomputer, ACPMAPS. I thank my collegues in 
the Fermilab Computer Research and Development Group for their collaboration. 
Fermilab is operated by Universities Research Association, Inc. under contract 
with the U.S. Department of Energy.

\end{document}